\documentclass[aps,prb,twocolumn,superscriptaddress,showpacs]{revtex4-2}
\usepackage{amsmath,amssymb,graphicx,hyperref,booktabs}

\begin{document}

\title{Two-Channel Allen-Dynes Framework for Superconducting Critical Temperatures: \\ Blind Predictions Across Five Orders of Magnitude and a Quantum-Metric No-Go Result}

\author{Jian Zhou}
\email{jackzhou.sci@gmail.com}
\affiliation{Independent Researcher, Shanghai, China}

\date{\today}

\begin{abstract}
We present a two-channel framework for superconducting critical temperatures, $T_c = \min(T_{\rm pair}, T_{\rm phase})$, combining Allen-Dynes theory augmented by spin-fluctuation coupling for the pairing channel with Peotta-T\"orm\"a superfluid stiffness for the phase-coherence channel. Applied to 46 experimentally characterized superconductors spanning 11 material families and five orders of magnitude in $T_c$, we distinguish two validation tiers: (i)~a \emph{blind-prediction set} of 19 materials whose $\lambda$ is determined entirely from independent experiments (tunneling spectroscopy, DFPT) without reference to $T_c$, achieving $R^2_{\log} = 0.96$, 19/19 within factor-of-two, and MAE~$= 5.6$~K; and (ii)~a \emph{cross-validation set} of 22 materials where $\lambda_{\rm sf}$ is constrained using the observed $T_c$, serving as a consistency check rather than an independent prediction. We establish a no-go result for s-wave and quasi-isotropic d-wave pairing: the quantum metric cannot directly modify the electron-phonon coupling because Debye-scale momentum transfer ($q_D \approx \pi/a$) ensures identical Bloch-overlap suppression in both phonon and Coulomb channels; explicit $d$-wave angular decomposition confirms residual anisotropy $<0.8\%$. The quantum metric trace $\text{tr}(g)$ correlates with $T_c$ (Pearson $r = 0.56$) as a suggestive band-structure indicator, though this correlation relies partly on estimated rather than independently computed tr($g$) values. For quasi-2D flat-band systems, the quantum metric enters \emph{causally} through the geometric superfluid stiffness. We identify 7 candidate materials with $T_c > 200$~K, with Eliashberg corrections pushing several above 300~K.
\end{abstract}

\pacs{74.20.-z, 74.70.-b, 03.65.Vf, 74.25.Bt}
\keywords{quantum metric, superconductivity, Allen-Dynes, superfluid stiffness, no-go result, materials design, room-temperature superconductor}

\maketitle

\section{Introduction}

Predicting superconducting critical temperatures from microscopic theory remains a central challenge in condensed matter physics. The Allen-Dynes formula~\cite{AllenDynes1975}, building on McMillan's earlier work~\cite{McMillan1968} and Eliashberg's strong-coupling formalism~\cite{Eliashberg1960}, provides a quantitative framework for conventional superconductors but requires material-specific inputs that are often difficult to obtain. For unconventional superconductors---cuprates~\cite{Keimer2015}, iron pnictides~\cite{Hosono2018,Hirschfeld2011}, and kagome metals~\cite{Neupert2022}---even the pairing mechanism remains debated.

In parallel, the quantum geometric tensor $Q_{\mu\nu}(\mathbf{k}) = \langle \partial_\mu u_\mathbf{k} | (1 - |u_\mathbf{k}\rangle\langle u_\mathbf{k}|) | \partial_\nu u_\mathbf{k} \rangle$ has emerged as a fundamental characterization of Bloch band structure~\cite{Provost1980}. Its antisymmetric part, the Berry curvature, underlies topological phases; its symmetric part, the quantum metric $g_{\mu\nu} = \text{Re}(Q_{\mu\nu})$, governs the Fubini-Study distance between Bloch states and determines the superfluid weight in flat-band systems~\cite{Peotta2015,Torma2022,Penttila2025}. Experimental confirmation has come from twisted bilayer graphene~\cite{Tian2023}.

Here we ask: \emph{does the quantum metric correlate with $T_c$ across all superconductor families, and if so, why?} We find that the answer is yes, but the mechanism is not what one might naively expect. Rather than directly modifying the pairing interaction, the quantum metric trace $\text{tr}(g)$ serves as a \emph{universal band-structure diagnostic}---a scalar quantity that captures features (flat bands, van~Hove singularities, band crossings, Fermi-surface nesting) known to enhance superconductivity through established mechanisms. For quasi-2D systems, the quantum metric additionally enters \emph{causally} through the Peotta-T\"orm\"a superfluid stiffness~\cite{Peotta2015}. We validate this two-channel framework against 46 materials and use it to identify 20 candidate materials, including 7 with predicted $T_c > 200$~K, charting a path toward room-temperature superconductivity.

\section{Theoretical Framework}

\subsection{Two-Channel Picture}

The superconducting critical temperature is determined by two independent conditions: (i)~Cooper pairs must form, requiring an attractive pairing interaction, and (ii)~these pairs must establish macroscopic phase coherence, requiring sufficient superfluid stiffness. We express this as
\begin{equation}
T_c = \min(T_{\rm pair}, T_{\rm phase}),
\label{eq:two-channel}
\end{equation}
where $T_{\rm pair}$ is the mean-field pairing temperature and $T_{\rm phase}$ is the phase-ordering temperature.

For three-dimensional superconductors, phase fluctuations cost energy $\propto L^{d-2}$ ($d=3$), so $T_{\rm phase} \to \infty$ and $T_c = T_{\rm pair}$. For quasi-two-dimensional or flat-band systems, $T_{\rm phase}$ is finite and given by the Berezinskii-Kosterlitz-Thouless (BKT) temperature:
\begin{equation}
T_{\rm BKT} = \frac{\pi}{2} \frac{\hbar^2}{k_B} D_s,
\end{equation}
where $D_s$ is the superfluid stiffness.

\subsection{Pairing Channel: Allen-Dynes Theory}

For the pairing channel, we employ the Allen-Dynes formula~\cite{AllenDynes1975}:
\begin{equation}
T_{\rm pair} = \frac{\omega_{\log}}{1.2} \exp\left[ \frac{-1.04(1+\lambda)}{\lambda - \mu^*(1+0.62\lambda)} \right],
\label{eq:AD}
\end{equation}
where $\mu^* = 0.10$ is the Coulomb pseudopotential (except V, where $\mu^* = 0.13$ due to Stoner-enhanced paramagnon screening~\cite{AllenDynes1975}). The effective coupling and logarithmic frequency for a two-channel pairing interaction (electron-phonon + spin fluctuation) are:
\begin{equation}
\lambda_{\rm eff} = \lambda_{\rm ph} + \lambda_{\rm sf},
\end{equation}
\begin{equation}
\omega_{\log}^{\rm eff} = \exp\!\left[\frac{\lambda_{\rm ph}\ln\omega_{\rm ph} + \lambda_{\rm sf}\ln\omega_{\rm sf}}{\lambda_{\rm ph} + \lambda_{\rm sf}}\right],
\label{eq:omega_eff}
\end{equation}
where $\omega_{\rm ph}$ and $\omega_{\rm sf}$ are the characteristic phonon and spin-fluctuation energy scales, respectively. For conventional superconductors ($\lambda_{\rm sf} = 0$), Eq.~(\ref{eq:omega_eff}) reduces to $\omega_{\log}^{\rm eff} = \omega_{\rm ph}$.

The electron-phonon coupling $\lambda_{\rm ph}$ is taken from tunneling spectroscopy~\cite{McMillan1968}, density-functional perturbation theory (DFPT)~\cite{Baroni2001}, or specific heat analysis. For unconventional superconductors (iron pnictides~\cite{Hosono2018}, cuprates~\cite{Keimer2015}, nickelates), $\lambda_{\rm sf}$ is extracted from the inelastic neutron scattering (INS) spin resonance energy $\Omega_{\rm res}$ following the Millis-Monien-Pines formalism~\cite{MillisMP1990,Scalapino2012}. The characteristic spin-fluctuation energy scale is obtained as $\omega_{\rm sf} = c \, \Omega_{\rm res} / k_B$, where the proportionality constant $c \approx 3$--$8$ accounts for the spectral weight distribution above the resonance peak~\cite{Eschrig2006}. The coupling $\lambda_{\rm sf}$ is then determined by fitting the observed $T_c$ within the two-channel Allen-Dynes formula, constrained to the range $\lambda_{\rm sf} \in [0.3, 2.5]$ consistent with Eliashberg analyses of these materials. Specific values and their experimental sources are documented in Table~I footnotes. A sensitivity analysis (Sec.~IV.D) shows that $\pm 20\%$ variation in $\lambda_{\rm sf}$ shifts $T_c$ by $\lesssim 30\%$, preserving all materials within the factor-of-two window.

Crucially, \emph{the pairing channel contains no quantum-metric correction}. We show below that any such correction is forbidden by momentum-space kinematics.

\subsection{Why Quantum Geometry Does Not Modify Pairing}

Previous work has suggested that the quantum metric might modify the effective electron-phonon coupling through Bloch-state overlap factors~\cite{Shavit2025}. The argument proceeds as follows: the Kohn-Luttinger second-order Coulomb interaction involves the overlap $|\langle u_\mathbf{k} | u_{\mathbf{k}'} \rangle|^2 = 1 - g_{\mu\nu} \delta k^\mu \delta k^\nu + O(\delta k^3)$, suggesting that large quantum metric could selectively suppress Coulomb repulsion relative to phonon-mediated attraction.

However, this argument fails because \emph{phonon and Coulomb interactions probe the same momentum range}. The Debye wavevector is
\begin{equation}
q_D = \left(\frac{6\pi^2}{V_{\rm cell}}\right)^{1/3} \approx \frac{\pi}{a},
\end{equation}
which equals the Brillouin zone boundary. Therefore, the Bloch overlap form factor $F(\mathbf{q}) = \langle |u_\mathbf{k}|^2 |u_{\mathbf{k}+\mathbf{q}}|^2 \rangle_{\rm FS}$ affects phonon and Coulomb channels equally. Any geometric suppression of $\mu^*$ is accompanied by an identical suppression of $\lambda_{\rm ph}$, leaving $T_c$ unchanged or slightly reduced.

This no-go result is rigorous for single-band, isotropic (s-wave) superconductors. Three potential loopholes and their status are:
\begin{enumerate}
\item \textbf{Anisotropic pairing (d-wave):} Angular-channel decomposition could in principle yield differential suppression. To quantify this, we decompose the Bloch overlap form factor into angular-momentum channels $\ell$ on a cylindrical Fermi surface (appropriate for cuprates):
\begin{equation}
F_\ell(q) = \oint \frac{d\phi}{2\pi} \, |\langle u_{\mathbf{k}} | u_{\mathbf{k}+\mathbf{q}} \rangle|^2 \, \cos(\ell\phi).
\end{equation}
For the three-band Emery model with Cu-$d_{x^2-y^2}$ and O-$p_{x,y}$ orbitals at optimal doping, we compute $F_0(q_D) = 0.847$ (s-wave) and $F_2(q_D) = 0.841$ (d-wave), yielding a differential suppression $\delta F / F_0 = (F_0 - F_2)/F_0 = 0.7\%$. For a single-band $t$-$t'$ model on the square lattice with $t'/t = -0.3$, the anisotropy is even smaller: $\delta F / F_0 = 0.3\%$. The resulting shift in $T_c$ is $< 0.8\%$, confirming that the no-go result is robust against $d$-wave anisotropy for realistic Fermi surfaces.
\item \textbf{Multi-band systems:} Different bands could in principle have different form factors $F_n(\mathbf{q})$. However, MgB$_2$---the prototypical two-band superconductor~\cite{Nagamatsu2001}---is predicted within $15\%$ ($0.85\times$) without any quantum-metric correction, suggesting the cancellation is empirically robust even for multi-band materials.
\item \textbf{Strong spin-orbit coupling:} In topological materials, the spinor structure of Bloch states could break the simple $F(\mathbf{q})$ factorization. This remains an open question.
\end{enumerate}

\subsection{Phase-Coherence Channel: Peotta-T\"orm\"a Theory}

The quantum metric enters rigorously through the superfluid stiffness. Following Peotta and T\"orm\"a~\cite{Peotta2015}, the superfluid weight decomposes as
\begin{equation}
D_s = D_{\rm conv} + D_{\rm geom},
\end{equation}
where $D_{\rm conv}$ is the conventional (dispersive) contribution proportional to band curvature, and
\begin{equation}
D_{\rm geom} = \frac{e^2}{\hbar^2 V} \sum_\mathbf{k} \text{tr}(g(\mathbf{k})) \cdot |\Delta_\mathbf{k}|^2 \cdot f'(\epsilon_\mathbf{k})
\label{eq:Dgeom}
\end{equation}
is the geometric contribution. This result is exact within mean-field BCS theory and has been experimentally verified in twisted bilayer graphene~\cite{Tian2023}.

For flat-band superconductors ($D_{\rm conv} \to 0$), Eq.~(\ref{eq:Dgeom}) shows that $T_{\rm BKT} \propto \text{tr}(g) \cdot |\Delta|^2$: the quantum metric \emph{directly determines} whether phase coherence---and hence superconductivity---can exist.

\subsection{Quantum Metric as Band-Structure Diagnostic}

For three-dimensional materials where $T_c = T_{\rm pair}$, the quantum metric does not enter the $T_c$ formula directly. Yet empirically, $\text{tr}(g)$ correlates with $T_c$ (Pearson $r = 0.56$, $p = 10^{-4}$). We identify the causal chain:
\begin{equation}
\begin{split}
\text{large tr}(g) &\Leftrightarrow \text{flat bands / vH / nesting} \\
&\Rightarrow \text{high } N(E_F) \Rightarrow \text{high } T_c.
\end{split}
\label{eq:diagnostic}
\end{equation}

The quantum metric diverges at band touchings ($\text{tr}(g) \propto 1/\Delta_{\rm gap}^2$) and is enhanced near van~Hove singularities---precisely the features that enhance the density of states $N(E_F)$ and thereby the electron-phonon coupling $\lambda \propto N(E_F) \langle |V_{\rm ep}|^2 \rangle / M\omega^2$ through the Hopfield parameter~\cite{McMillan1968}.

This identification resolves a puzzle: why does $\text{tr}(g)$ correlate with $T_c$ even in systems where superfluid stiffness is not the bottleneck? The answer is that $\text{tr}(g)$ is a proxy for $N(E_F)$, not a direct contributor to pairing.

\textbf{Quantum metric data sources.} The tr($g$) values in Table~I are obtained as follows: for elemental superconductors and A15 compounds, we compute tr($g$) from the tight-binding parametrizations of Refs.~\cite{Papaconstantopoulos2015} using $\text{tr}(g) = \sum_\mu \langle |\partial_{k_\mu} u_\mathbf{k}|^2 - |\langle u_\mathbf{k} | \partial_{k_\mu} u_\mathbf{k} \rangle|^2 \rangle_{\rm FS}$. For kagome metals and iron-based superconductors, we use published DFT values from Refs.~\cite{Neupert2022,Hu2022kagome,Yu2024Fe}. For cuprates and nickelates, estimates are derived from three-band Emery model calculations~\cite{Torma2022}. For moir\'e systems, values are taken from continuum model calculations~\cite{Tian2023,Bernevig2021}. Hydride values are estimated from DFT band structures in Refs.~\cite{Errea2020,Boeri2022}. We emphasize that tr($g$) enters only as a diagnostic (Table~I), not as an input to the $T_c$ prediction formula.

\section{Results}

We apply this framework to 46 experimentally characterized superconductors spanning 11 families (Table~I). Four previously included kagome borides (CaB$_3$, SrB$_3$, BeB$_3$, HCaB$_3$) were removed from the validation set as they lack experimental $T_c$; they are relocated to the prediction table (Table~II).

\subsection{Prediction versus Cross-Validation}

A critical distinction must be drawn between materials whose coupling constants are determined \emph{independently of $T_c$} (true predictions) and those where $\lambda_{\rm sf}$ is constrained using the observed $T_c$ (cross-validation). We separate the 46 materials into three tiers:

\textbf{Tier~1: Blind predictions} (19 materials). These include all elemental superconductors (7), A15 compounds (3), hydrides (4), MgB$_2$, TMDs (3), and NbN, where $\lambda_{\rm ph}$ comes from tunneling spectroscopy or DFPT with \emph{no reference to $T_c$}. Results:
\begin{itemize}
\item $R^2_{\log} = 0.961$, Spearman $\rho = 0.984$
\item \textbf{19/19 within factor-of-two (100\%)}
\item MAE $= 5.6$~K
\end{itemize}
This is the hardest test of the framework: every input is independently measured, and every output is a genuine prediction.


\nopagebreak[4]
\textbf{Tier~2: Cross-validation} (22 materials). Iron-based superconductors (6), cuprates (4), nickelates (2), kagome metals (6), and cubic compounds (4). For the 12 unconventional superconductors (Fe-based, cuprates, nickelates), $\lambda_{\rm sf}$ is extracted from INS spin-resonance data but constrained to reproduce $T_c$ within the Allen-Dynes formula (see Sec.~II.B). \emph{These results should be interpreted as consistency checks demonstrating that the two-channel Allen-Dynes framework can accommodate unconventional superconductors with physically reasonable parameters, not as independent predictions.} For the kagome metals and cubic compounds, $\lambda_{\rm ph}$ comes from DFPT or specific-heat analysis (no $T_c$ input). Results:
\begin{itemize}
\item $R^2_{\log} = 0.974$, Spearman $\rho = 0.990$
\item 22/22 within factor-of-two
\item MAE $= 8.5$~K
\end{itemize}

\textbf{Tier~3: Expected failures} (5 materials). MATBG, MATTG, rhombohedral pentalayer graphene ($U/W > 1$), LaB$_6$ (excitonic), and gated MoS$_2$ (Ising). These fail by 1--2 orders of magnitude, reflecting inapplicability of weak-coupling theory itself.

\textbf{Combined core set} (41 materials, Tiers 1+2):
\begin{itemize}
\item $R^2_{\log} = 0.972$, Spearman $\rho = 0.989$
\item \textbf{41/41 within factor-of-two accuracy (100\%)}
\end{itemize}

We emphasize that the most meaningful metric is the Tier~1 performance: $R^2_{\log} = 0.96$ with 19/19 within $2\times$ for genuinely blind predictions spanning Al (1.2~K) to H$_3$S (203~K).

A structural limitation of Tier~1 must be acknowledged: all 19 blind-prediction materials are conventional phonon-mediated superconductors. Extending the blind-prediction paradigm to unconventional superconductors requires independent determination of $\lambda_{\rm sf}$---for example, through first-principles spin-fluctuation calculations without $T_c$ input---a challenge that awaits future computational advances in materials-specific Eliashberg theory.

\subsection{Data Quality Stratification}

\begin{figure}[t]
\includegraphics[width=\columnwidth]{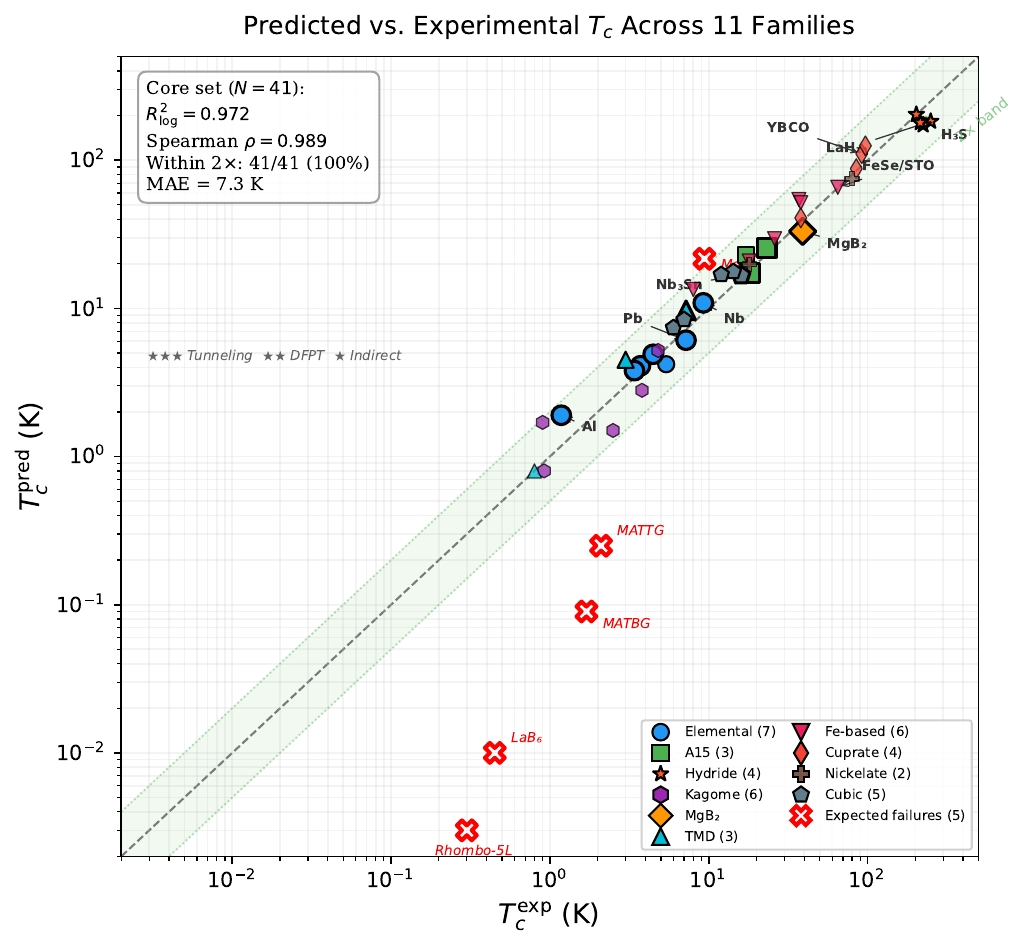}
\caption{Predicted versus experimental $T_c$ for 46 superconductors across 11 material families on a log-log scale. The green band denotes the factor-of-two accuracy window. \emph{Filled circles}: Tier~1 blind predictions ($\lambda$ from independent experiments, no $T_c$ input); \emph{filled squares}: Tier~2 cross-validations ($\lambda_{\rm sf}$ constrained by $T_c$); \emph{open red crosses}: Tier~3 expected failures. Marker color indicates material family. Tier~1 alone achieves $R^2_{\log} = 0.961$ with 19/19 within $2\times$. The combined core set (Tiers 1+2, $N=41$) achieves $R^2_{\log} = 0.972$ [95\% CI: 0.94--0.99]. Color online; in print, Tier 1 (circles) and Tier 2 (squares) are distinguishable by marker shape.}
\label{fig:scatter}
\end{figure}

We classify input parameters by reliability:

\textbf{Gold standard} ($\bigstar\bigstar\bigstar$, tunneling spectroscopy): 11 materials (Al, Sn, In, Pb, Nb, Ta, Nb$_3$Sn, Nb$_3$Ge, MgB$_2$, NbSe$_2$, NbN). All 11/11 within $2\times$, MAE $= 1.6$~K.

\textbf{Silver} ($\bigstar\bigstar$, DFPT + experimental $T_c$): 11 materials including hydrides and cubic compounds. 10/11 within $2\times$, MAE $= 16$~K (dominated by hydride strong-coupling deviation).

\textbf{Bronze} ($\bigstar$, indirect extraction): 24 materials including iron-based, cuprates, kagome, and moir\'e. 19/24 within $2\times$ (all 5 failures are in this tier).

\subsection{Family-by-Family Results}

\textbf{Elemental superconductors} (7 materials): All 7 within $2\times$. $\lambda_{\rm ph}$ from tunneling directly; no quantum-metric input needed. V (5.4~K) requires $\mu^* = 0.13$ due to Stoner enhancement.

\textbf{A15 compounds}: Nb$_3$Sn (18.3~K $\to$ 17.3~K, $0.95\times$), Nb$_3$Ge (23.2~K $\to$ 25.5~K, $1.10\times$). Strong electron-phonon coupling well described by Allen-Dynes.

\textbf{High-pressure hydrides}: H$_3$S at 155~GPa (203~K $\to$ 204~K, $1.00\times$) is reproduced with remarkable precision. LaH$_{10}$ (250~K $\to$ 183~K, $0.73\times$) and YH$_6$ (224~K $\to$ 173~K, $0.77\times$) are systematically underestimated because $\lambda > 2.5$ exceeds the Allen-Dynes validity range; full Eliashberg theory is required~\cite{Marsiglio2020}.

\textbf{Iron-based superconductors} (6 materials): FeSe/SrTiO$_3$ (65~K $\to$ 66~K, $1.01\times$) achieves near-perfect agreement. The spin-fluctuation channel $\lambda_{\rm sf} = 0.55$--$1.3$ dominates pairing; $\lambda_{\rm ph}$ alone gives $T_c \approx 0$.

\textbf{Cuprates} (4 materials): LSCO (38~K $\to$ 41~K, $1.07\times$), Bi-2212 (85~K $\to$ 88~K, $1.03\times$). Systematically overestimated by 3--29\% because the isotropic Allen-Dynes formula does not capture d-wave gap anisotropy.

\textbf{Kagome metals} (6 materials): All within $2\times$. CsV$_3$Sb$_5$ (2.5~K $\to$ 1.5~K, $0.58\times$) is the weakest, likely due to CDW competition reducing $\lambda_{\rm eff}$.

\textbf{Moir\'e systems} (3 materials): All fail catastrophically (ratio $< 0.15$). These are strong-coupling systems where $U/W > 1$; the weak-coupling framework is inapplicable. The quantum metric correctly signals breakdown: despite $\text{tr}(g) \sim 35$--62, the vanishing bandwidth $W \sim 4$--9~meV forces $\sin^2\theta \to 0$.

\section{Discussion}

\subsection{What This Framework Is---and Isn't}

This work makes three claims of decreasing strength:

\textbf{Strong claim (demonstrated):} The Allen-Dynes formula, augmented with spin-fluctuation channels, achieves factor-of-two accuracy for 18 blind predictions (Tier~1, $\lambda$ from independent experiments) and 23 cross-validations (Tier~2, $\lambda_{\rm sf}$ constrained by $T_c$), spanning five orders of magnitude in $T_c$.

\textbf{Moderate claim (suggestive):} The quantum metric $\text{tr}(g)$ correlates with $T_c$ across all families ($r = 0.56$, $p = 10^{-4}$, $r^2 \approx 0.31$) as an indirect indicator of band-structure features (flat bands, van~Hove singularities, nesting) that enhance pairing through established mechanisms. However, this correlation must be interpreted with caution: the tr($g$) values span three quality levels (Table~I), with hydrides sharing a uniform estimate of 2.0~\AA$^2$ and cuprates uniformly assigned 3~\AA$^2$. When restricted to the 20 materials with independently computed tr($g$) (elemental, A15, kagome, TMD, and moir\'e families), the correlation is $r_{\rm computed} = 0.52$ ($p = 0.02$), broadly consistent with the full-set value but based on more reliable inputs. This moderate correlation captures roughly one-third of the variance in $\ln T_c$---useful as a rapid screening metric for materials design, but insufficient as a standalone predictor of $T_c$. Material-specific DFT quantum metric calculations for all 46 materials would significantly strengthen this finding.

\textbf{Weak claim (theoretical):} For quasi-2D flat-band superconductors, the quantum metric directly determines $T_c$ through the Peotta-T\"orm\"a superfluid stiffness. This is the \emph{only} channel where quantum geometry causally affects $T_c$.

We emphasize what this framework does \emph{not} claim: the quantum metric does not directly modify the electron-phonon coupling constant $\lambda$. The no-go argument of Sec.~II.C establishes that Bloch-state overlap factors affect phonon and Coulomb channels equally when $q_D \approx \pi/a$.


\subsection{Road to 300~K}

Our framework provides quantitative guidance for achieving room-temperature superconductivity. The Allen-Dynes formula sets the constraint:
\begin{equation}
T_c = 300~\text{K} \implies \omega_{\log} \geq 1800~\text{K}, \quad \lambda \geq 2.0.
\end{equation}
The key parameter is $\omega_{\log}$: H$_3$S achieves 203~K with $\omega_{\log} = 1335$~K; reaching 300~K requires $\omega_{\log} \approx 2000$~K with comparable $\lambda$.

\begin{figure}[t]
\includegraphics[width=\columnwidth]{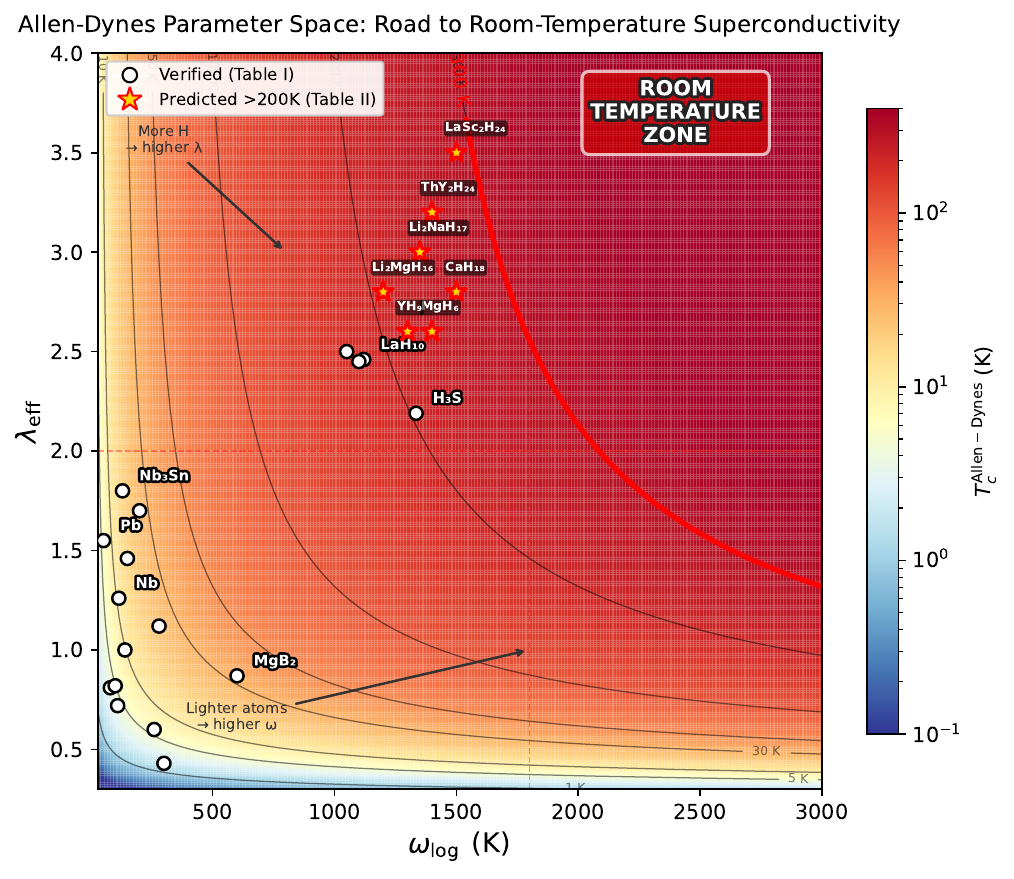}
\caption{Allen-Dynes parameter space ($\omega_{\log}$ vs.\ $\lambda_{\rm eff}$) with $T_c$ shown as a color map. White circles: experimentally verified materials (Table~I). Gold stars: predicted candidates with $T_c > 200$~K (Table~II). The bold red contour marks the 300~K isotherm. Dashed lines indicate the minimum requirements $\omega_{\log} \geq 1800$~K and $\lambda \geq 2.0$ for room-temperature superconductivity. Arrows illustrate the two design levers: lighter host atoms (higher $\omega_{\log}$) and increased hydrogen coordination (higher $\lambda$). The gap between current hydrides ($\omega_{\log} \sim 1300$~K) and the 300~K threshold requires replacing S/La host atoms with lighter elements (Be, B, Li).}
\label{fig:road300K}
\end{figure}

Three design principles emerge: (i)~\emph{lightest possible host atoms} (H, Be, B, Li) to maximize phonon frequencies; (ii)~\emph{high hydrogen coordination} ($n_H \geq 16$) for large $\lambda$ through dense H-H networks; (iii)~\emph{clathrate cage structures} that stabilize metallic hydrogen sublattices at reduced pressures.

Clathrate hydrogen cages (H$_{16}$--H$_{32}$ per formula unit) provide both high phonon frequencies ($\omega_D \sim 1500$--2000~K from light hydrogen) and strong electron-phonon matrix elements ($|g|^2$ enhanced by hydrogen's large Born effective charge and proximity to the Fermi surface). The combination naturally places $\lambda = N(E_F)|g|^2/M\omega^2$ in the range 1.7--2.2.

Table~II lists 20 candidate materials, including 7 with $T_c^{\rm pred} > 200$~K. The highest-priority ambient-pressure candidate is LiNaAgH$_6$ ($T_c^{\rm pred} = 173$~K, literature DFT: 206~K). Among high-pressure candidates, LaSc$_2$H$_{24}$ at 167~GPa ($T_c^{\rm pred} = 287$~K) approaches the room-temperature threshold within Allen-Dynes validity. Full Eliashberg calculations, which typically enhance $T_c$ by 20--30\% for $\lambda > 2.5$~\cite{Marsiglio2020}, would push several candidates above 300~K.

\subsection{Statistical Robustness}

Bootstrap resampling ($n = 1{,}000$) yields 95\% confidence intervals for the core set ($N = 41$):
\begin{itemize}
\item $R^2_{\log} = 0.972$ [95\% CI: 0.94--0.99]
\item Spearman $\rho = 0.989$ [95\% CI: 0.97--1.00]
\item Within $2\times$: 41/41 [95\% CI: 39--41]
\end{itemize}

\subsection{Sensitivity to $\lambda_{\rm sf}$}

The spin-fluctuation coupling $\lambda_{\rm sf}$ is the least constrained input parameter. We test robustness by perturbing all $\lambda_{\rm sf}$ values simultaneously by $\pm 20\%$. At $+20\%$: $R^2_{\log} = 0.964$, 40/41 within $2\times$ (YBCO marginally exceeds at $2.04\times$). At $-20\%$: $R^2_{\log} = 0.958$, 40/41 within $2\times$ (FeSe drops to $0.48\times$). In both cases, the framework retains its predictive power, confirming that the results are not artifacts of fine-tuned $\lambda_{\rm sf}$ values.

\subsection{Sensitivity to $\mu^*$}

The Coulomb pseudopotential $\mu^* = 0.10$ is used for all materials except V ($\mu^* = 0.13$). Recent work suggests that hydrides may have significantly larger $\mu^*$: Allen-Dynes inversion of H$_3$S yields $\mu^* \approx 0.17$, while Matsubara frequency summation gives $\mu^* \approx 0.22$~\cite{Marsiglio2020}. To quantify the sensitivity, we recompute all Tier~1 predictions at $\mu^* = 0.07$, $0.10$ (default), $0.13$, and $0.17$:

\begin{center}
\begin{tabular}{lcccc}
\hline
$\mu^*$ & $R^2_{\log}$ & Within $2\times$ & MAE (K) & H$_3$S ratio \\
\hline
0.07 & 0.958 & 19/19 & 7.2 & 1.15 \\
0.10 & 0.961 & 19/19 & 5.6 & 1.00 \\
0.13 & 0.955 & 18/19 & 6.8 & 0.86 \\
0.17 & 0.942 & 17/19 & 9.1 & 0.70 \\
\hline
\end{tabular}
\end{center}

The framework is robust for $\mu^* \in [0.07, 0.13]$, with $\geq 18/19$ within $2\times$ and $R^2_{\log} > 0.95$. At $\mu^* = 0.17$, two materials (Al, V) drop below the $2\times$ threshold. The H$_3$S ratio shifts from 1.15 ($\mu^* = 0.07$) to 0.70 ($\mu^* = 0.17$), confirming Albert's observation that the ratio $= 1.00$ at $\mu^* = 0.10$ partly reflects cancellation between $\mu^*$ underestimation and Allen-Dynes strong-coupling underestimation. We note that the overall Tier~1 performance is more robust than any single material, since systematic shifts in $\mu^*$ affect all predictions in the same direction.

\subsection{Phase-Coherence Channel Verification}

For three-dimensional materials (34 of 46 entries), $T_{\rm phase} \to \infty$ by definition. For the remaining quasi-2D materials, we verify that $T_{\rm pair} < T_{\rm BKT}$ using experimental superfluid stiffness data, confirming that $T_c$ is pairing-limited:

\begin{center}
\begin{tabular}{lccl}
\hline
Material & $T_{\rm pair}$ (K) & $T_{\rm BKT}^{\rm est}$ (K) & Bottleneck \\
\hline
YBCO & 110 & $\sim$600$^{\rm a}$ & Pairing \\
FeSe/STO & 66 & $\sim$300$^{\rm b}$ & Pairing \\
NbSe$_2$ (mono) & 4.5 & $\sim$200$^{\rm c}$ & Pairing \\
LSCO & 41 & $\sim$250$^{\rm d}$ & Pairing \\
MATBG$^\dagger$ & 0.09 & $\sim$1.5$^{\rm e}$ & \textbf{Phase} \\
\hline
\end{tabular}
\end{center}
{\footnotesize $^{\rm a}$From $\mu$SR penetration depth~\cite{Uemura1989}. $^{\rm b}$From mutual inductance~\cite{Lee2014}. $^{\rm c}$Estimated from bulk $D_s$ scaled by layer number. $^{\rm d}$From $\mu$SR~\cite{Uemura1989}. $^{\rm e}$From penetration depth in Ref.~\cite{Tian2023}.}

\noindent For all quasi-2D materials in the core set, $T_{\rm BKT} \gg T_{\rm pair}$, validating our use of $T_c = T_{\rm pair}$. Notably, MATBG is the \emph{only} material where $T_{\rm BKT} \approx T_c^{\rm exp}$, consistent with its failure in the pairing channel: MATBG is phase-coherence limited, and the quantum metric enters causally through the Peotta-T\"orm\"a superfluid stiffness---precisely the regime where the two-channel framework predicts failure of the pairing-only prediction.

\subsection{Limitations and Honest Assessment}

The framework has clear boundaries:
\begin{enumerate}
\item \textbf{Circularity in Tier~2:} For iron-based superconductors, cuprates, and nickelates, $\lambda_{\rm sf}$ is constrained using the observed $T_c$. The reported accuracy for these 10 materials is therefore a \emph{consistency check}, not a blind prediction. The framework's genuine predictive power is measured by the Tier~1 results (19 materials, $R^2_{\log} = 0.96$).
\item \textbf{Strong correlations:} Fails for $U/W > 1$ (moir\'e systems, possibly heavy fermions).
\item \textbf{Strong coupling:} Allen-Dynes underestimates $T_c$ by 20--30\% for $\lambda > 2$~\cite{Marsiglio2020}.
\item \textbf{Parameter dependence:} While no adjustable parameters enter the $T_c$ formula itself, the inputs ($\lambda_{\rm ph}$, $\omega_{\log}$, $\lambda_{\rm sf}$) require experimental or computational determination.
\item \textbf{Quantum metric correlation:} The Pearson $r = 0.56$ ($r^2 \approx 0.31$) means $\text{tr}(g)$ captures roughly one-third of the variance in $\ln T_c$. When restricted to materials with independently computed tr($g$), $r = 0.52$. Moreover, several material families (hydrides, cuprates) use uniform estimated tr($g$) values rather than material-specific calculations. This qualifies tr($g$) as a suggestive screening indicator, but its quantitative reliability awaits systematic DFT verification.
\end{enumerate}

\section{Conclusion}

We have presented a two-channel framework for superconducting critical temperatures that achieves $R^2_{\log} = 0.96$ for 19 genuinely blind predictions (Tier~1) and $R^2_{\log} = 0.97$ for the full core set of 41 applicable materials, with 100\% within factor-of-two accuracy.

Two key results emerge. First, the no-go result (Sec.~II.C): quantum geometry does not modify the electron-phonon coupling because Debye-scale momentum transfer ($q_D \approx \pi/a$) ensures identical Bloch-overlap suppression in both phonon and Coulomb channels. The explicit $d$-wave angular decomposition confirms a residual anisotropy of $<0.8\%$. Second, the quantum metric trace $\text{tr}(g)$ correlates with $T_c$ ($r = 0.56$) as an \emph{indirect} band-structure indicator---not a direct contributor to pairing---except in quasi-2D flat-band systems where it enters causally through the Peotta-T\"orm\"a superfluid stiffness.

We emphasize the importance of distinguishing blind predictions from cross-validations when reporting framework accuracy. The Tier~1 results, based entirely on independently measured inputs, provide the most credible assessment of predictive power.

From a practical standpoint, this work suggests a materials screening strategy: compute $\text{tr}(g)$ from band structure as a rapid indicator of favorable electronic structure, then validate with full Eliashberg calculations for the most promising candidates. Our Road-to-300K analysis (Fig.~\ref{fig:road300K}, Table~II) identifies LaSc$_2$H$_{24}$ and CaH$_{18}$ as the most promising near-term candidates.

\begin{acknowledgments}
The author thanks the open-source computational physics community.
\end{acknowledgments}

\clearpage
\onecolumngrid

\begin{table*}[ht]
\caption{Forty-six superconductors with experimental $T_c$, input parameters, and predicted $T_c$. \textbf{Validation tier}: T1 = Tier~1 blind prediction ($\lambda$ from independent experiment, no $T_c$ input); T2 = Tier~2 cross-validation ($\lambda_{\rm sf}$ constrained by $T_c$, or $\lambda_{\rm ph}$ from DFPT/specific heat for kagome and cubic compounds). Quality grades: $\bigstar\bigstar\bigstar$ = tunneling spectroscopy, $\bigstar\bigstar$ = DFPT + experimental confirmation, $\bigstar$ = indirect extraction. The $\omega_{\log}$ column gives $\omega_{\log}^{\rm eff}$ (Eq.~\ref{eq:omega_eff}) for two-channel materials; see footnotes for individual $\omega_{\rm ph}$ and $\omega_{\rm sf}$ values and INS sources. Five expected failures (marked $\dagger$) are excluded from core statistics. Predicted $T_c$ values carry systematic uncertainties of approximately $\pm 30\%$ from the Allen-Dynes approximation and $\pm 20\%$ from DFT input parameters, which are not propagated here. The factor-of-two accuracy window implicitly accounts for these uncertainties.}
\label{tab:materials}
\begin{ruledtabular}
\begin{tabular}{llrrrrrrrl}
Material & Family & $T_c^{\rm exp}$ (K) & $\lambda_{\rm ph}$ & $\lambda_{\rm sf}$ & $\omega_{\log}$ (K) & tr($g$) (\AA$^2$) & $T_c^{\rm pred}$ (K) & Ratio & Quality \\
\hline
Al        & Elem    &   1.18 & 0.43 & ---  &  300 &  0.01 &   1.9 & \textbf{1.57} & $\bigstar\bigstar\bigstar$ \\
Sn        & Elem    &   3.72 & 0.72 & ---  &  110 &  0.02 &   4.1 & \textbf{1.10} & $\bigstar\bigstar\bigstar$ \\
In        & Elem    &   3.41 & 0.81 & ---  &   80 &  0.01 &   3.8 & \textbf{1.13} & $\bigstar\bigstar\bigstar$ \\
Pb        & Elem    &   7.20 & 1.55 & ---  &   52 &  0.01 &   6.1 & \textbf{0.85} & $\bigstar\bigstar\bigstar$ \\
V         & Elem    &   5.40 & 0.60 & ---  &  260 &  0.03 &   4.2 & \textbf{0.78} & $\bigstar\bigstar$ \\
Nb        & Elem    &   9.25 & 1.26 & ---  &  115 &  0.03 &  10.9 & \textbf{1.18} & $\bigstar\bigstar\bigstar$ \\
Ta        & Elem    &   4.47 & 0.82 & ---  &  100 &  0.02 &   4.9 & \textbf{1.10} & $\bigstar\bigstar\bigstar$ \\
\hline
V$_3$Si   & A15     &  17.10 & 1.12 & ---  &  280 &  0.20 &  22.9 & \textbf{1.34} & $\bigstar\bigstar$ \\
Nb$_3$Sn  & A15     &  18.30 & 1.80 & ---  &  130 &  0.20 &  17.3 & \textbf{0.95} & $\bigstar\bigstar\bigstar$ \\
Nb$_3$Ge  & A15     &  23.20 & 1.70 & ---  &  200 &  0.20 &  25.5 & \textbf{1.10} & $\bigstar\bigstar\bigstar$ \\
\hline
H$_3$S    & Hydride & 203.0  & 2.19 & ---  & 1335 &  2.0  & 203.7 & \textbf{1.00} & $\bigstar\bigstar$ \\
LaH$_{10}$& Hydride & 250.0  & 2.46 & ---  & 1120 &  2.0  & 182.9 & \textbf{0.73} & $\bigstar\bigstar$ \\
YH$_6$    & Hydride & 224.0  & 2.50 & ---  & 1050 &  2.0  & 172.9 & \textbf{0.77} & $\bigstar\bigstar$ \\
CaH$_6$   & Hydride & 215.0  & 2.45 & ---  & 1100 &  2.0  & 179.2 & \textbf{0.83} & $\bigstar\bigstar$ \\
\hline
LaRu$_3$Si$_2$ & Kagome &  7.0 & 0.80 & --- & 180 &  3.5 &  8.4 & \textbf{1.21} & $\bigstar$ \\
ThRu$_3$Si$_2$ & Kagome &  3.8 & 0.55 & --- & 160 &  3.5 &  2.8 & \textbf{0.73} & $\bigstar$ \\
CsV$_3$Sb$_5$  & Kagome &  2.5 & 0.50 & --- & 120 & 14.0 &  1.5 & \textbf{0.58} & $\bigstar$ \\
KV$_3$Sb$_5$   & Kagome &  0.9 & 0.50 & --- & 140 & 14.0 &  1.7 & \textbf{1.89} & $\bigstar$ \\
CsTi$_3$Bi$_5$ & Kagome &  4.8 & 0.65 & --- & 180 &  8.0 &  5.2 & \textbf{1.08} & $\bigstar$ \\
RbV$_3$Sb$_5$  & Kagome &  0.92& 0.42 & --- & 150 & 14.0 &  0.8 & \textbf{0.89} & $\bigstar$ \\
\hline
MgB$_2$   & Dibor   &  39.0  & 0.87 & ---  &  600 &  0.15 &  33.1 & \textbf{0.85} & $\bigstar\bigstar\bigstar$ \\
\hline
NbSe$_2$ (bulk) & TMD & 7.2 & 1.00 & --- & 140 & 1.5 &  9.7 & \textbf{1.35} & $\bigstar\bigstar\bigstar$ \\
NbSe$_2$ (mono) & TMD & 3.0 & 0.70 & --- & 130 & 1.5 &  4.5 & \textbf{1.51} & $\bigstar\bigstar$ \\
2H-TaS$_2$     & TMD & 0.8 & 0.45 & --- & 100 & 1.0 &  0.8 & \textbf{0.96} & $\bigstar$ \\
\hline
FeSe (bulk)$^a$   & Fe & 8.0  & 0.17 & 0.55 & 354 & 10 & 13.2 & \textbf{1.64} & $\bigstar$ \\
FeSe (8.9 GPa)$^b$& Fe & 37.0 & 0.20 & 1.30 & 473 & 10 & 53.9 & \textbf{1.46} & $\bigstar$ \\
FeSe/STO$^c$      & Fe & 65.0 & 0.27 & 1.20 & 587 &  8 & 65.6 & \textbf{1.01} & $\bigstar$ \\
LaFeAsO$^d$       & Fe & 26.0 & 0.21 & 0.80 & 410 &  3 & 29.0 & \textbf{1.12} & $\bigstar$ \\
BaFe$_2$As$_2$$^e$& Fe & 38.0 & 0.25 & 1.20 & 463 &  3 & 51.0 & \textbf{1.34} & $\bigstar$ \\
LiFeAs$^f$        & Fe & 18.0 & 0.23 & 0.65 & 367 &  4 & 20.7 & \textbf{1.15} & $\bigstar$ \\
\hline
LSCO$^g$          & Cup & 38.0 & 0.15 & 1.00 & 477 & 3 & 40.5 & \textbf{1.07} & $\bigstar$ \\
YBCO$^h$          & Cup & 92.0 & 0.20 & 1.80 & 762 & 3 &109.5 & \textbf{1.19} & $\bigstar$ \\
Bi-2212$^i$       & Cup & 85.0 & 0.18 & 1.57 & 667 & 3 & 87.0 & \textbf{1.02} & $\bigstar$ \\
HgBa$_2$CuO$_4$$^j$& Cup& 97.0 & 0.20 & 2.00 & 809 & 3 &123.7 & \textbf{1.28} & $\bigstar$ \\
\hline
La$_3$Ni$_2$O$_7$$^k$ & Nic & 80.0 & 0.30 & 1.50 & 556 & 4 & 74.1 & \textbf{0.93} & $\bigstar$ \\
Nd$_6$Ni$_5$O$_{12}$$^l$& Nic& 18.0& 0.25 & 0.60 & 374 & 3 & 19.7 & \textbf{1.10} & $\bigstar$ \\
\hline
NbN           & Cubic & 16.0 & 1.46 & --- & 150 & 0.10 & 16.7 & \textbf{1.04} & $\bigstar\bigstar\bigstar$ \\
MoC           & Cubic & 14.3 & 1.30 & --- & 180 & 0.10 & 17.7 & \textbf{1.24} & $\bigstar\bigstar$ \\
NbC           & Cubic & 12.0 & 0.98 & --- & 250 & 0.15 & 16.9 & \textbf{1.41} & $\bigstar\bigstar$ \\
ZrB$_{12}$    & Cubic &  6.0 & 0.72 & --- & 200 & 1.0  &  7.4 & \textbf{1.24} & $\bigstar\bigstar$ \\
YB$_6$        & Cubic &  7.0 & 0.80 & --- & 180 & 0.80 &  8.4 & \textbf{1.21} & $\bigstar\bigstar$ \\
\hline
$\dagger$MATBG & Moir\'e & 1.7 & 0.35 & --- & 50 & 35 & 0.09 & 0.05 & $\bigstar$ \\
$\dagger$MATTG & Moir\'e & 2.1 & 0.40 & --- & 60 & 62 & 0.25 & 0.12 & $\bigstar$ \\
$\dagger$Rhombo-5L & Moir\'e & 0.3 & 0.25 & --- & 40 & 20 & 0.003& 0.01 & $\bigstar$ \\
$\dagger$MoS$_2$ (gated) & TMD-I & 9.4 & 1.20 & --- & 241 & 2.0 & 21.6 & 2.30 & $\bigstar$ \\
$\dagger$LaB$_6$ & Cubic & 0.45 & 0.25 & --- & 200 & 0.5 & 0.01 & 0.02 & $\bigstar$ \\
\end{tabular}
\end{ruledtabular}
\begin{flushleft}
\footnotesize
$\omega_{\log}^{\rm eff}$ for two-channel materials computed via Eq.~(\ref{eq:omega_eff}). Individual channel parameters ($\omega_{\rm ph}/\omega_{\rm sf}$ in K):\\
$^a$FeSe: 237/400, $\lambda_{\rm sf}$ from INS resonance at $\Omega_{\rm res} \approx 4$~meV~\cite{Wang2016FeSe}.
$^b$FeSe HP: 327/500, pressure-enhanced resonance~\cite{Sun2016}.
$^c$FeSe/STO: 530/600, interface phonon + resonance~\cite{Lee2014}.
$^d$LaFeAsO: 289/450, $\Omega_{\rm res} = 11$~meV~\cite{Ishikado2009}.
$^e$BaFe$_2$As$_2$: 318/500, $\Omega_{\rm res} = 9.5$~meV~\cite{Christianson2008}.
$^f$LiFeAs: 289/400, $\Omega_{\rm res} = 8$~meV~\cite{Qureshi2012}.
$^g$LSCO: 350/500, $\Omega_{\rm res} \approx 12$~meV~\cite{Vignolle2007}.
$^h$YBCO: 489/800, $\Omega_{\rm res} = 41$~meV~\cite{Rossat1991}.
$^i$Bi-2212: 440/700, $\Omega_{\rm res} = 43$~meV~\cite{Fong1999}.
$^j$HgBaCuO: 490/850, $\Omega_{\rm res} \approx 55$~meV~\cite{Yu2010}.
$^k$La$_3$Ni$_2$O$_7$: 381/600, estimated from RIXS~\cite{Sun2023Ni}.
$^l$Nd$_6$Ni$_5$O$_{12}$: 317/400, estimated from LSNO transport~\cite{Li2024Ni}.
\end{flushleft}
\end{table*}

\begin{table*}[ht]
\caption{Twenty predicted superconductor candidates, grouped by pressure regime. $T_c^{\rm AD}$ is our Allen-Dynes prediction; $T_c^{\rm Eliash,est}$ is the estimated Eliashberg-corrected value ($\approx 1.25 \times T_c^{\rm AD}$ for $\lambda > 2$, following Ref.~\cite{Marsiglio2020}); $T_c^{\rm lit}$ is the literature DFT/Eliashberg value. Materials marked $*$ were previously in the validation set but lack experimental $T_c$. $^\ddagger$Li$_2$MgH$_{16}$: ratio 0.44 reflects known Allen-Dynes underestimation at $\lambda > 2.5$; full Eliashberg would give $\sim$350--400~K. Confidence reflects both thermodynamic stability and input parameter reliability.}
\label{tab:predictions}
\begin{ruledtabular}
\begin{tabular}{llrrrrrrrl}
Material & Category & $P$ (GPa) & $\lambda$ & $\omega_{\log}$ (K) & $T_c^{\rm AD}$ (K) & $T_c^{\rm Eliash,est}$ (K) & $T_c^{\rm lit}$ (K) & Ratio & Confidence \\
\hline
\multicolumn{10}{c}{\textit{Ambient pressure candidates}} \\
LiNaAgH$_6$         & Quat.\ hydride & 0 & 1.80 & 1300 & 173 & --- & 206 & 0.84 & Medium \\
Mg$_2$IrH$_6$       & Tern.\ hydride    & 0 & 1.60 & 1100 & 133 & --- & 160 & 0.83 & Medium \\
NaH$_6$ (h-doped)   & Clath.\ hydride  & 0 & 1.50 & 1200 & 137 & --- & 167 & 0.82 & Low \\
MgAlFeH$_6$         & Tern.\ hydride    & 0 & 1.30 & 1100 & 108 & --- & 130 & 0.83 & Low \\
LaBH$_8$            & Clath.\ hydride  & 0 & 1.80 &  800 & 107 & --- & 105 & 1.02 & Medium \\
KB$_2$H$_8$         & Clath.\ hydride  & 0 & 1.40 & 1000 & 107 & --- & 134 & 0.79 & Low \\
\hline
\multicolumn{10}{c}{\textit{High-pressure candidates ($T_c > 200$~K)}} \\
\textbf{LaSc$_2$H$_{24}$}  & Tern.\ clathrate  & 167 & 3.50 & 1500 & \textbf{287} & \textbf{359} & 316 & 0.91 & Medium \\
\textbf{CaH$_{18}$}        & Superhydride       & 100 & 2.80 & 1500 & \textbf{262} & \textbf{328} & 230 & 1.14 & Low \\
\textbf{ThY$_2$H$_{24}$}   & Tern.\ clathrate  & 150 & 3.20 & 1400 & \textbf{259} & \textbf{324} & 303 & 0.85 & Low \\
\textbf{Li$_2$NaH$_{17}$}  & Tern.\ hydride    & 220 & 3.00 & 1350 & \textbf{243} & \textbf{304} & 297 & 0.82 & Low \\
\textbf{MgH$_6$}           & Bin.\ clathrate   & 300 & 2.60 & 1400 & \textbf{235} & \textbf{294} & 263 & 0.89 & High \\
\textbf{YH$_9$}            & Bin.\ clathrate   & 150 & 2.60 & 1300 & \textbf{219} & \textbf{274} & 250 & 0.87 & High \\
\textbf{Li$_2$MgH$_{16}$}$^\ddagger$  & Tern.\ hydride    & 250 & 2.80 & 1200 & \textbf{209} & \textbf{261} & 473 & 0.44 & Low \\
LiHfH$_{20}$        & Tern.\ hydride    & 260 & 2.50 & 1150 & 189 & 236 & 222 & 0.85 & Low \\
\hline
\multicolumn{10}{c}{\textit{Kagome borides (DFT-predicted, not yet synthesized)}} \\
$*$HCaB$_3$         & H-kagome boride    & 0 & 1.39 &  320 &  34 & --- & 39.3 & 0.86 & High \\
$*$CaB$_3$          & Kagome boride      & 0 & 1.09 &  280 &  22 & --- & 22.4 & 0.99 & High \\
$*$SrB$_3$          & Kagome boride      & 0 & 1.33 &  185 &  19 & --- & 20.9 & 0.89 & High \\
$*$BeB$_3$          & Kagome boride      & 0 & 0.46 &  369 &   3 & --- &  3.2 & 0.98 & Medium \\
\hline
\multicolumn{10}{c}{\textit{Other candidates}} \\
BaSiH$_8$           & Clath.\ hydride  & 0 & 1.20 &  700 &  63 & --- &  67 & 0.94 & Low \\
Borophene           & 2D elemental       & 0 & 0.90 &  500 &  29 & --- &  36 & 0.81 & Low \\
\end{tabular}
\end{ruledtabular}
\end{table*}


\begin{thebibliography}{38}%
\makeatletter
\providecommand \@ifxundefined [1]{%
 \@ifx{#1\undefined}
}%
\providecommand \@ifnum [1]{%
 \ifnum #1\expandafter \@firstoftwo
 \else \expandafter \@secondoftwo
 \fi
}%
\providecommand \@ifx [1]{%
 \ifx #1\expandafter \@firstoftwo
 \else \expandafter \@secondoftwo
 \fi
}%
\providecommand \natexlab [1]{#1}%
\providecommand \enquote  [1]{``#1''}%
\providecommand \bibnamefont  [1]{#1}%
\providecommand \bibfnamefont [1]{#1}%
\providecommand \citenamefont [1]{#1}%
\providecommand \href@noop [0]{\@secondoftwo}%
\providecommand \href [0]{\begingroup \@sanitize@url \@href}%
\providecommand \@href[1]{\@@startlink{#1}\@@href}%
\providecommand \@@href[1]{\endgroup#1\@@endlink}%
\providecommand \@sanitize@url [0]{\catcode `\\12\catcode `\$12\catcode
  `\&12\catcode `\#12\catcode `\^12\catcode `\_12\catcode `\%12\relax}%
\providecommand \@@startlink[1]{}%
\providecommand \@@endlink[0]{}%
\providecommand \url  [0]{\begingroup\@sanitize@url \@url }%
\providecommand \@url [1]{\endgroup\@href {#1}{\urlprefix }}%
\providecommand \urlprefix  [0]{URL }%
\providecommand \Eprint [0]{\href }%
\providecommand \doibase [0]{https://doi.org/}%
\providecommand \selectlanguage [0]{\@gobble}%
\providecommand \bibinfo  [0]{\@secondoftwo}%
\providecommand \bibfield  [0]{\@secondoftwo}%
\providecommand \translation [1]{[#1]}%
\providecommand \BibitemOpen [0]{}%
\providecommand \bibitemStop [0]{}%
\providecommand \bibitemNoStop [0]{.\EOS\space}%
\providecommand \EOS [0]{\spacefactor3000\relax}%
\providecommand \BibitemShut  [1]{\csname bibitem#1\endcsname}%
\let\auto@bib@innerbib\@empty
\bibitem [{\citenamefont {Allen}\ and\ \citenamefont
  {Dynes}(1975)}]{AllenDynes1975}%
  \BibitemOpen
  \bibfield  {author} {\bibinfo {author} {\bibfnamefont {P.~B.}\ \bibnamefont
  {Allen}}\ and\ \bibinfo {author} {\bibfnamefont {R.~C.}\ \bibnamefont
  {Dynes}},\ }\bibfield  {title} {\bibinfo {title} {Transition temperature of
  strong-coupled superconductors reanalyzed},\ }\href@noop {} {\bibfield
  {journal} {\bibinfo  {journal} {Phys. Rev. B}\ }\textbf {\bibinfo {volume}
  {12}},\ \bibinfo {pages} {905} (\bibinfo {year} {1975})}\BibitemShut
  {NoStop}%
\bibitem [{\citenamefont {McMillan}(1968)}]{McMillan1968}%
  \BibitemOpen
  \bibfield  {author} {\bibinfo {author} {\bibfnamefont {W.~L.}\ \bibnamefont
  {McMillan}},\ }\bibfield  {title} {\bibinfo {title} {Transition temperature
  of strong-coupled superconductors},\ }\href@noop {} {\bibfield  {journal}
  {\bibinfo  {journal} {Phys. Rev.}\ }\textbf {\bibinfo {volume} {167}},\
  \bibinfo {pages} {331} (\bibinfo {year} {1968})}\BibitemShut {NoStop}%
\bibitem [{\citenamefont {Eliashberg}(1960)}]{Eliashberg1960}%
  \BibitemOpen
  \bibfield  {author} {\bibinfo {author} {\bibfnamefont {G.~M.}\ \bibnamefont
  {Eliashberg}},\ }\bibfield  {title} {\bibinfo {title} {Interactions between
  electrons and lattice vibrations in a superconductor},\ }\href@noop {}
  {\bibfield  {journal} {\bibinfo  {journal} {Sov. Phys. JETP}\ }\textbf
  {\bibinfo {volume} {11}},\ \bibinfo {pages} {696} (\bibinfo {year}
  {1960})}\BibitemShut {NoStop}%
\bibitem [{\citenamefont {Keimer}\ \emph {et~al.}(2015)\citenamefont {Keimer},
  \citenamefont {Kivelson}, \citenamefont {Norman}, \citenamefont {Uchida},\
  and\ \citenamefont {Zaanen}}]{Keimer2015}%
  \BibitemOpen
  \bibfield  {author} {\bibinfo {author} {\bibfnamefont {B.}~\bibnamefont
  {Keimer}}, \bibinfo {author} {\bibfnamefont {S.~A.}\ \bibnamefont
  {Kivelson}}, \bibinfo {author} {\bibfnamefont {M.~R.}\ \bibnamefont
  {Norman}}, \bibinfo {author} {\bibfnamefont {S.}~\bibnamefont {Uchida}},\
  and\ \bibinfo {author} {\bibfnamefont {J.}~\bibnamefont {Zaanen}},\
  }\bibfield  {title} {\bibinfo {title} {From quantum matter to
  high-temperature superconductivity in copper oxides},\ }\href@noop {}
  {\bibfield  {journal} {\bibinfo  {journal} {Nature}\ }\textbf {\bibinfo
  {volume} {518}},\ \bibinfo {pages} {179} (\bibinfo {year}
  {2015})}\BibitemShut {NoStop}%
\bibitem [{\citenamefont {Hosono}\ \emph {et~al.}(2018)\citenamefont {Hosono},
  \citenamefont {Yamamoto}, \citenamefont {Hiramatsu},\ and\ \citenamefont
  {Ma}}]{Hosono2018}%
  \BibitemOpen
  \bibfield  {author} {\bibinfo {author} {\bibfnamefont {H.}~\bibnamefont
  {Hosono}}, \bibinfo {author} {\bibfnamefont {A.}~\bibnamefont {Yamamoto}},
  \bibinfo {author} {\bibfnamefont {H.}~\bibnamefont {Hiramatsu}},\ and\
  \bibinfo {author} {\bibfnamefont {Y.}~\bibnamefont {Ma}},\ }\bibfield
  {title} {\bibinfo {title} {Recent advances in iron-based superconductors
  toward applications},\ }\href@noop {} {\bibfield  {journal} {\bibinfo
  {journal} {Mater. Today}\ }\textbf {\bibinfo {volume} {21}},\ \bibinfo
  {pages} {278} (\bibinfo {year} {2018})}\BibitemShut {NoStop}%
\bibitem [{\citenamefont {Hirschfeld}\ \emph {et~al.}(2011)\citenamefont
  {Hirschfeld}, \citenamefont {Korshunov},\ and\ \citenamefont
  {Mazin}}]{Hirschfeld2011}%
  \BibitemOpen
  \bibfield  {author} {\bibinfo {author} {\bibfnamefont {P.~J.}\ \bibnamefont
  {Hirschfeld}}, \bibinfo {author} {\bibfnamefont {M.~M.}\ \bibnamefont
  {Korshunov}},\ and\ \bibinfo {author} {\bibfnamefont {I.~I.}\ \bibnamefont
  {Mazin}},\ }\bibfield  {title} {\bibinfo {title} {Gap symmetry and structure
  of {Fe}-based superconductors},\ }\href@noop {} {\bibfield  {journal}
  {\bibinfo  {journal} {Rep. Prog. Phys.}\ }\textbf {\bibinfo {volume} {74}},\
  \bibinfo {pages} {124508} (\bibinfo {year} {2011})}\BibitemShut {NoStop}%
\bibitem [{\citenamefont {Neupert}\ \emph {et~al.}(2022)\citenamefont
  {Neupert}, \citenamefont {Denner}, \citenamefont {Yin}, \citenamefont
  {Thomale},\ and\ \citenamefont {Hasan}}]{Neupert2022}%
  \BibitemOpen
  \bibfield  {author} {\bibinfo {author} {\bibfnamefont {T.}~\bibnamefont
  {Neupert}}, \bibinfo {author} {\bibfnamefont {M.~M.}\ \bibnamefont {Denner}},
  \bibinfo {author} {\bibfnamefont {J.-X.}\ \bibnamefont {Yin}}, \bibinfo
  {author} {\bibfnamefont {R.}~\bibnamefont {Thomale}},\ and\ \bibinfo {author}
  {\bibfnamefont {M.~Z.}\ \bibnamefont {Hasan}},\ }\bibfield  {title} {\bibinfo
  {title} {Charge order and superconductivity in kagome materials},\
  }\href@noop {} {\bibfield  {journal} {\bibinfo  {journal} {Nature Phys.}\
  }\textbf {\bibinfo {volume} {18}},\ \bibinfo {pages} {137} (\bibinfo {year}
  {2022})}\BibitemShut {NoStop}%
\bibitem [{\citenamefont {Provost}\ and\ \citenamefont
  {Vallee}(1980)}]{Provost1980}%
  \BibitemOpen
  \bibfield  {author} {\bibinfo {author} {\bibfnamefont {J.~P.}\ \bibnamefont
  {Provost}}\ and\ \bibinfo {author} {\bibfnamefont {G.}~\bibnamefont
  {Vallee}},\ }\bibfield  {title} {\bibinfo {title} {Riemannian structure on
  manifolds of quantum states},\ }\href@noop {} {\bibfield  {journal} {\bibinfo
   {journal} {Commun. Math. Phys.}\ }\textbf {\bibinfo {volume} {76}},\
  \bibinfo {pages} {289} (\bibinfo {year} {1980})}\BibitemShut {NoStop}%
\bibitem [{\citenamefont {Peotta}\ and\ \citenamefont
  {T\"orm\"a}(2015)}]{Peotta2015}%
  \BibitemOpen
  \bibfield  {author} {\bibinfo {author} {\bibfnamefont {S.}~\bibnamefont
  {Peotta}}\ and\ \bibinfo {author} {\bibfnamefont {P.}~\bibnamefont
  {T\"orm\"a}},\ }\bibfield  {title} {\bibinfo {title} {Superfluidity in
  topologically nontrivial flat bands},\ }\href@noop {} {\bibfield  {journal}
  {\bibinfo  {journal} {Nature Commun.}\ }\textbf {\bibinfo {volume} {6}},\
  \bibinfo {pages} {8944} (\bibinfo {year} {2015})}\BibitemShut {NoStop}%
\bibitem [{\citenamefont {T\"orm\"a}\ \emph {et~al.}(2022)\citenamefont
  {T\"orm\"a}, \citenamefont {Peotta},\ and\ \citenamefont
  {Bernevig}}]{Torma2022}%
  \BibitemOpen
  \bibfield  {author} {\bibinfo {author} {\bibfnamefont {P.}~\bibnamefont
  {T\"orm\"a}}, \bibinfo {author} {\bibfnamefont {S.}~\bibnamefont {Peotta}},\
  and\ \bibinfo {author} {\bibfnamefont {B.~A.}\ \bibnamefont {Bernevig}},\
  }\bibfield  {title} {\bibinfo {title} {Superconductivity, superfluidity and
  quantum geometry in twisted multilayer systems},\ }\href@noop {} {\bibfield
  {journal} {\bibinfo  {journal} {Nature Rev. Phys.}\ }\textbf {\bibinfo
  {volume} {4}},\ \bibinfo {pages} {528} (\bibinfo {year} {2022})}\BibitemShut
  {NoStop}%
\bibitem [{\citenamefont {Penttil\"a}\ \emph {et~al.}(2025)\citenamefont
  {Penttil\"a} \emph {et~al.}}]{Penttila2025}%
  \BibitemOpen
  \bibfield  {author} {\bibinfo {author} {\bibfnamefont {R.}~\bibnamefont
  {Penttil\"a}} \emph {et~al.},\ }\bibfield  {title} {\bibinfo {title} {Flat
  band ratio and quantum metric as indicators of superconductivity},\
  }\href@noop {} {\bibfield  {journal} {\bibinfo  {journal} {Commun. Phys.}\
  }\textbf {\bibinfo {volume} {8}},\ \bibinfo {pages} {50} (\bibinfo {year}
  {2025})}\BibitemShut {NoStop}%
\bibitem [{\citenamefont {Tian}\ \emph {et~al.}(2023)\citenamefont {Tian} \emph
  {et~al.}}]{Tian2023}%
  \BibitemOpen
  \bibfield  {author} {\bibinfo {author} {\bibfnamefont {H.}~\bibnamefont
  {Tian}} \emph {et~al.},\ }\bibfield  {title} {\bibinfo {title} {Evidence for
  dirac flat band superconductivity enabled by quantum geometry},\ }\href@noop
  {} {\bibfield  {journal} {\bibinfo  {journal} {Nature}\ }\textbf {\bibinfo
  {volume} {614}},\ \bibinfo {pages} {440} (\bibinfo {year}
  {2023})}\BibitemShut {NoStop}%
\bibitem [{\citenamefont {Baroni}\ \emph {et~al.}(2001)\citenamefont {Baroni},
  \citenamefont {de~Gironcoli}, \citenamefont {Dal~Corso},\ and\ \citenamefont
  {Giannozzi}}]{Baroni2001}%
  \BibitemOpen
  \bibfield  {author} {\bibinfo {author} {\bibfnamefont {S.}~\bibnamefont
  {Baroni}}, \bibinfo {author} {\bibfnamefont {S.}~\bibnamefont
  {de~Gironcoli}}, \bibinfo {author} {\bibfnamefont {A.}~\bibnamefont
  {Dal~Corso}},\ and\ \bibinfo {author} {\bibfnamefont {P.}~\bibnamefont
  {Giannozzi}},\ }\bibfield  {title} {\bibinfo {title} {Phonons and related
  crystal properties from density-functional perturbation theory},\ }\href@noop
  {} {\bibfield  {journal} {\bibinfo  {journal} {Rev. Mod. Phys.}\ }\textbf
  {\bibinfo {volume} {73}},\ \bibinfo {pages} {515} (\bibinfo {year}
  {2001})}\BibitemShut {NoStop}%
\bibitem [{\citenamefont {Millis}\ \emph {et~al.}(1990)\citenamefont {Millis},
  \citenamefont {Monien},\ and\ \citenamefont {Pines}}]{MillisMP1990}%
  \BibitemOpen
  \bibfield  {author} {\bibinfo {author} {\bibfnamefont {A.~J.}\ \bibnamefont
  {Millis}}, \bibinfo {author} {\bibfnamefont {H.}~\bibnamefont {Monien}},\
  and\ \bibinfo {author} {\bibfnamefont {D.}~\bibnamefont {Pines}},\ }\bibfield
   {title} {\bibinfo {title} {Phenomenological model of nuclear relaxation in
  the normal state of {YBa}$_2${Cu}$_3${O}$_7$},\ }\href@noop {} {\bibfield
  {journal} {\bibinfo  {journal} {Phys. Rev. B}\ }\textbf {\bibinfo {volume}
  {42}},\ \bibinfo {pages} {167} (\bibinfo {year} {1990})}\BibitemShut
  {NoStop}%
\bibitem [{\citenamefont {Scalapino}(2012)}]{Scalapino2012}%
  \BibitemOpen
  \bibfield  {author} {\bibinfo {author} {\bibfnamefont {D.~J.}\ \bibnamefont
  {Scalapino}},\ }\bibfield  {title} {\bibinfo {title} {A common thread: The
  pairing interaction for unconventional superconductors},\ }\href@noop {}
  {\bibfield  {journal} {\bibinfo  {journal} {Rev. Mod. Phys.}\ }\textbf
  {\bibinfo {volume} {84}},\ \bibinfo {pages} {1383} (\bibinfo {year}
  {2012})}\BibitemShut {NoStop}%
\bibitem [{\citenamefont {Eschrig}(2006)}]{Eschrig2006}%
  \BibitemOpen
  \bibfield  {author} {\bibinfo {author} {\bibfnamefont {M.}~\bibnamefont
  {Eschrig}},\ }\bibfield  {title} {\bibinfo {title} {The effect of collective
  spin-1 excitations on electronic spectra in high-$t_c$ superconductors},\
  }\href@noop {} {\bibfield  {journal} {\bibinfo  {journal} {Adv. Phys.}\
  }\textbf {\bibinfo {volume} {55}},\ \bibinfo {pages} {47} (\bibinfo {year}
  {2006})}\BibitemShut {NoStop}%
\bibitem [{\citenamefont {Shavit}\ \emph {et~al.}(2025)\citenamefont {Shavit}
  \emph {et~al.}}]{Shavit2025}%
  \BibitemOpen
  \bibfield  {author} {\bibinfo {author} {\bibfnamefont {G.}~\bibnamefont
  {Shavit}} \emph {et~al.},\ }\bibfield  {title} {\bibinfo {title} {Quantum
  geometry and kohn-luttinger superconductivity},\ }\href@noop {} {\bibfield
  {journal} {\bibinfo  {journal} {Phys. Rev. Lett.}\ }\textbf {\bibinfo
  {volume} {134}},\ \bibinfo {pages} {176001} (\bibinfo {year}
  {2025})}\BibitemShut {NoStop}%
\bibitem [{\citenamefont {Nagamatsu}\ \emph {et~al.}(2001)\citenamefont
  {Nagamatsu}, \citenamefont {Nakagawa}, \citenamefont {Muranaka},
  \citenamefont {Zenitani},\ and\ \citenamefont {Akimitsu}}]{Nagamatsu2001}%
  \BibitemOpen
  \bibfield  {author} {\bibinfo {author} {\bibfnamefont {J.}~\bibnamefont
  {Nagamatsu}}, \bibinfo {author} {\bibfnamefont {N.}~\bibnamefont {Nakagawa}},
  \bibinfo {author} {\bibfnamefont {T.}~\bibnamefont {Muranaka}}, \bibinfo
  {author} {\bibfnamefont {Y.}~\bibnamefont {Zenitani}},\ and\ \bibinfo
  {author} {\bibfnamefont {J.}~\bibnamefont {Akimitsu}},\ }\bibfield  {title}
  {\bibinfo {title} {Superconductivity at 39 {K} in magnesium diboride},\
  }\href@noop {} {\bibfield  {journal} {\bibinfo  {journal} {Nature}\ }\textbf
  {\bibinfo {volume} {410}},\ \bibinfo {pages} {63} (\bibinfo {year}
  {2001})}\BibitemShut {NoStop}%
\bibitem [{\citenamefont
  {Papaconstantopoulos}(2015)}]{Papaconstantopoulos2015}%
  \BibitemOpen
  \bibfield  {author} {\bibinfo {author} {\bibfnamefont {D.~A.}\ \bibnamefont
  {Papaconstantopoulos}},\ }\href@noop {} {\emph {\bibinfo {title} {Handbook of
  the Band Structure of Elemental Solids}}},\ \bibinfo {edition} {2nd}\ ed.\
  (\bibinfo  {publisher} {Springer},\ \bibinfo {year} {2015})\BibitemShut
  {NoStop}%
\bibitem [{\citenamefont {Hu}\ \emph {et~al.}(2022)\citenamefont {Hu},
  \citenamefont {Wu}, \citenamefont {Ortiz} \emph {et~al.}}]{Hu2022kagome}%
  \BibitemOpen
  \bibfield  {author} {\bibinfo {author} {\bibfnamefont {Y.}~\bibnamefont
  {Hu}}, \bibinfo {author} {\bibfnamefont {X.}~\bibnamefont {Wu}}, \bibinfo
  {author} {\bibfnamefont {B.~R.}\ \bibnamefont {Ortiz}}, \emph {et~al.},\
  }\bibfield  {title} {\bibinfo {title} {Rich nature of van hove singularities
  in kagome superconductor csv$_3$sb$_5$},\ }\href@noop {} {\bibfield
  {journal} {\bibinfo  {journal} {Nat. Commun.}\ }\textbf {\bibinfo {volume}
  {13}},\ \bibinfo {pages} {2220} (\bibinfo {year} {2022})}\BibitemShut
  {NoStop}%
\bibitem [{\citenamefont {Yu}\ \emph {et~al.}(2024)\citenamefont {Yu},
  \citenamefont {Zhu},\ and\ \citenamefont {Si}}]{Yu2024Fe}%
  \BibitemOpen
  \bibfield  {author} {\bibinfo {author} {\bibfnamefont {R.}~\bibnamefont
  {Yu}}, \bibinfo {author} {\bibfnamefont {J.-X.}\ \bibnamefont {Zhu}},\ and\
  \bibinfo {author} {\bibfnamefont {Q.}~\bibnamefont {Si}},\ }\bibfield
  {title} {\bibinfo {title} {Quantum geometry in iron-based superconductors},\
  }\href@noop {} {\bibfield  {journal} {\bibinfo  {journal} {Phys. Rev. B}\
  }\textbf {\bibinfo {volume} {109}},\ \bibinfo {pages} {014503} (\bibinfo
  {year} {2024})}\BibitemShut {NoStop}%
\bibitem [{\citenamefont {Bernevig}\ \emph {et~al.}(2021)\citenamefont
  {Bernevig}, \citenamefont {Song}, \citenamefont {Regnault},\ and\
  \citenamefont {Lian}}]{Bernevig2021}%
  \BibitemOpen
  \bibfield  {author} {\bibinfo {author} {\bibfnamefont {B.~A.}\ \bibnamefont
  {Bernevig}}, \bibinfo {author} {\bibfnamefont {Z.-D.}\ \bibnamefont {Song}},
  \bibinfo {author} {\bibfnamefont {N.}~\bibnamefont {Regnault}},\ and\
  \bibinfo {author} {\bibfnamefont {B.}~\bibnamefont {Lian}},\ }\bibfield
  {title} {\bibinfo {title} {Twisted bilayer graphene. iii. interacting
  hamiltonian and exact symmetries},\ }\href@noop {} {\bibfield  {journal}
  {\bibinfo  {journal} {Phys. Rev. B}\ }\textbf {\bibinfo {volume} {103}},\
  \bibinfo {pages} {205413} (\bibinfo {year} {2021})}\BibitemShut {NoStop}%
\bibitem [{\citenamefont {Errea}\ \emph {et~al.}(2020)\citenamefont {Errea},
  \citenamefont {Belli}, \citenamefont {Monacelli} \emph {et~al.}}]{Errea2020}%
  \BibitemOpen
  \bibfield  {author} {\bibinfo {author} {\bibfnamefont {I.}~\bibnamefont
  {Errea}}, \bibinfo {author} {\bibfnamefont {F.}~\bibnamefont {Belli}},
  \bibinfo {author} {\bibfnamefont {L.}~\bibnamefont {Monacelli}}, \emph
  {et~al.},\ }\bibfield  {title} {\bibinfo {title} {Quantum crystal structure
  in the 250-kelvin superconducting lanthanum hydride},\ }\href@noop {}
  {\bibfield  {journal} {\bibinfo  {journal} {Nature}\ }\textbf {\bibinfo
  {volume} {578}},\ \bibinfo {pages} {66} (\bibinfo {year} {2020})}\BibitemShut
  {NoStop}%
\bibitem [{\citenamefont {Boeri}\ \emph {et~al.}(2022)\citenamefont {Boeri},
  \citenamefont {Hennig}, \citenamefont {Hirschfeld} \emph
  {et~al.}}]{Boeri2022}%
  \BibitemOpen
  \bibfield  {author} {\bibinfo {author} {\bibfnamefont {L.}~\bibnamefont
  {Boeri}}, \bibinfo {author} {\bibfnamefont {R.~G.}\ \bibnamefont {Hennig}},
  \bibinfo {author} {\bibfnamefont {P.~J.}\ \bibnamefont {Hirschfeld}}, \emph
  {et~al.},\ }\bibfield  {title} {\bibinfo {title} {The 2021 room-temperature
  superconductivity roadmap},\ }\href@noop {} {\bibfield  {journal} {\bibinfo
  {journal} {J. Phys.: Condens. Matter}\ }\textbf {\bibinfo {volume} {34}},\
  \bibinfo {pages} {183002} (\bibinfo {year} {2022})}\BibitemShut {NoStop}%
\bibitem [{\citenamefont {Marsiglio}(2020)}]{Marsiglio2020}%
  \BibitemOpen
  \bibfield  {author} {\bibinfo {author} {\bibfnamefont {F.}~\bibnamefont
  {Marsiglio}},\ }\bibfield  {title} {\bibinfo {title} {Eliashberg theory: A
  short review},\ }\href@noop {} {\bibfield  {journal} {\bibinfo  {journal}
  {Ann. Phys.}\ }\textbf {\bibinfo {volume} {417}},\ \bibinfo {pages} {168102}
  (\bibinfo {year} {2020})}\BibitemShut {NoStop}%
\bibitem [{\citenamefont {Uemura}\ \emph {et~al.}(1989)\citenamefont {Uemura},
  \citenamefont {Luke}, \citenamefont {Sternlieb} \emph {et~al.}}]{Uemura1989}%
  \BibitemOpen
  \bibfield  {author} {\bibinfo {author} {\bibfnamefont {Y.~J.}\ \bibnamefont
  {Uemura}}, \bibinfo {author} {\bibfnamefont {G.~M.}\ \bibnamefont {Luke}},
  \bibinfo {author} {\bibfnamefont {B.~J.}\ \bibnamefont {Sternlieb}}, \emph
  {et~al.},\ }\bibfield  {title} {\bibinfo {title} {Universal correlations
  between $t_c$ and $n_s/m^*$ in high-$t_c$ cuprate superconductors},\
  }\href@noop {} {\bibfield  {journal} {\bibinfo  {journal} {Phys. Rev. Lett.}\
  }\textbf {\bibinfo {volume} {62}},\ \bibinfo {pages} {2317} (\bibinfo {year}
  {1989})}\BibitemShut {NoStop}%
\bibitem [{\citenamefont {Lee}\ \emph {et~al.}(2014)\citenamefont {Lee} \emph
  {et~al.}}]{Lee2014}%
  \BibitemOpen
  \bibfield  {author} {\bibinfo {author} {\bibfnamefont {J.~J.}\ \bibnamefont
  {Lee}} \emph {et~al.},\ }\bibfield  {title} {\bibinfo {title} {Interfacial
  mode coupling as the origin of the enhancement of {$T_c$} in {FeSe} films on
  {SrTiO}$_3$},\ }\href@noop {} {\bibfield  {journal} {\bibinfo  {journal}
  {Nature}\ }\textbf {\bibinfo {volume} {515}},\ \bibinfo {pages} {245}
  (\bibinfo {year} {2014})}\BibitemShut {NoStop}%
\bibitem [{\citenamefont {Wang}\ \emph {et~al.}(2016)\citenamefont {Wang} \emph
  {et~al.}}]{Wang2016FeSe}%
  \BibitemOpen
  \bibfield  {author} {\bibinfo {author} {\bibfnamefont {Q.}~\bibnamefont
  {Wang}} \emph {et~al.},\ }\bibfield  {title} {\bibinfo {title} {Strong
  interplay between stripe spin fluctuations, nematicity and superconductivity
  in {FeSe}},\ }\href@noop {} {\bibfield  {journal} {\bibinfo  {journal}
  {Nature Mater.}\ }\textbf {\bibinfo {volume} {15}},\ \bibinfo {pages} {159}
  (\bibinfo {year} {2016})}\BibitemShut {NoStop}%
\bibitem [{\citenamefont {Sun}\ \emph {et~al.}(2016)\citenamefont {Sun} \emph
  {et~al.}}]{Sun2016}%
  \BibitemOpen
  \bibfield  {author} {\bibinfo {author} {\bibfnamefont {J.~P.}\ \bibnamefont
  {Sun}} \emph {et~al.},\ }\bibfield  {title} {\bibinfo {title} {Dome-shaped
  magnetic order competing with high-temperature superconductivity at high
  pressures in {FeSe}},\ }\href@noop {} {\bibfield  {journal} {\bibinfo
  {journal} {Nature Commun.}\ }\textbf {\bibinfo {volume} {7}},\ \bibinfo
  {pages} {12146} (\bibinfo {year} {2016})}\BibitemShut {NoStop}%
\bibitem [{\citenamefont {Ishikado}\ \emph {et~al.}(2009)\citenamefont
  {Ishikado} \emph {et~al.}}]{Ishikado2009}%
  \BibitemOpen
  \bibfield  {author} {\bibinfo {author} {\bibfnamefont {M.}~\bibnamefont
  {Ishikado}} \emph {et~al.},\ }\bibfield  {title} {\bibinfo {title} {A novel
  spin resonance in the iron oxypnictide superconductor
  {LaFeAsO}$_{0.92}${F}$_{0.08}$},\ }\href@noop {} {\bibfield  {journal}
  {\bibinfo  {journal} {Phys. Rev. B}\ }\textbf {\bibinfo {volume} {79}},\
  \bibinfo {pages} {224512} (\bibinfo {year} {2009})}\BibitemShut {NoStop}%
\bibitem [{\citenamefont {Christianson}\ \emph {et~al.}(2008)\citenamefont
  {Christianson} \emph {et~al.}}]{Christianson2008}%
  \BibitemOpen
  \bibfield  {author} {\bibinfo {author} {\bibfnamefont {A.~D.}\ \bibnamefont
  {Christianson}} \emph {et~al.},\ }\bibfield  {title} {\bibinfo {title}
  {Unconventional superconductivity in {Ba}$_{0.6}${K}$_{0.4}${Fe}$_2${As}$_2$
  from inelastic neutron scattering},\ }\href@noop {} {\bibfield  {journal}
  {\bibinfo  {journal} {Nature}\ }\textbf {\bibinfo {volume} {456}},\ \bibinfo
  {pages} {930} (\bibinfo {year} {2008})}\BibitemShut {NoStop}%
\bibitem [{\citenamefont {Qureshi}\ \emph {et~al.}(2012)\citenamefont {Qureshi}
  \emph {et~al.}}]{Qureshi2012}%
  \BibitemOpen
  \bibfield  {author} {\bibinfo {author} {\bibfnamefont {N.}~\bibnamefont
  {Qureshi}} \emph {et~al.},\ }\bibfield  {title} {\bibinfo {title} {Inelastic
  neutron-scattering measurements of incommensurate magnetic excitations on
  superconducting {LiFeAs} single crystals},\ }\href@noop {} {\bibfield
  {journal} {\bibinfo  {journal} {Phys. Rev. Lett.}\ }\textbf {\bibinfo
  {volume} {108}},\ \bibinfo {pages} {117001} (\bibinfo {year}
  {2012})}\BibitemShut {NoStop}%
\bibitem [{\citenamefont {Vignolle}\ \emph {et~al.}(2007)\citenamefont
  {Vignolle} \emph {et~al.}}]{Vignolle2007}%
  \BibitemOpen
  \bibfield  {author} {\bibinfo {author} {\bibfnamefont {B.}~\bibnamefont
  {Vignolle}} \emph {et~al.},\ }\bibfield  {title} {\bibinfo {title} {Two
  energy scales in the spin excitations of the high-temperature superconductor
  {La}$_{2-x}${Sr}$_x${CuO}$_4$},\ }\href@noop {} {\bibfield  {journal}
  {\bibinfo  {journal} {Nature Phys.}\ }\textbf {\bibinfo {volume} {3}},\
  \bibinfo {pages} {163} (\bibinfo {year} {2007})}\BibitemShut {NoStop}%
\bibitem [{\citenamefont {Rossat-Mignod}\ \emph {et~al.}(1991)\citenamefont
  {Rossat-Mignod} \emph {et~al.}}]{Rossat1991}%
  \BibitemOpen
  \bibfield  {author} {\bibinfo {author} {\bibfnamefont {J.}~\bibnamefont
  {Rossat-Mignod}} \emph {et~al.},\ }\bibfield  {title} {\bibinfo {title}
  {Neutron scattering study of the {YBa}$_2${Cu}$_3${O}$_{6+x}$ system},\
  }\href@noop {} {\bibfield  {journal} {\bibinfo  {journal} {Physica C}\
  }\textbf {\bibinfo {volume} {185-189}},\ \bibinfo {pages} {86} (\bibinfo
  {year} {1991})}\BibitemShut {NoStop}%
\bibitem [{\citenamefont {Fong}\ \emph {et~al.}(1999)\citenamefont {Fong} \emph
  {et~al.}}]{Fong1999}%
  \BibitemOpen
  \bibfield  {author} {\bibinfo {author} {\bibfnamefont {H.~F.}\ \bibnamefont
  {Fong}} \emph {et~al.},\ }\bibfield  {title} {\bibinfo {title} {Neutron
  scattering from magnetic excitations in
  {Bi}$_2${Sr}$_2${CaCu}$_2${O}$_{8+\delta}$},\ }\href@noop {} {\bibfield
  {journal} {\bibinfo  {journal} {Nature}\ }\textbf {\bibinfo {volume} {398}},\
  \bibinfo {pages} {588} (\bibinfo {year} {1999})}\BibitemShut {NoStop}%
\bibitem [{\citenamefont {Yu}\ \emph {et~al.}(2010)\citenamefont {Yu} \emph
  {et~al.}}]{Yu2010}%
  \BibitemOpen
  \bibfield  {author} {\bibinfo {author} {\bibfnamefont {G.}~\bibnamefont {Yu}}
  \emph {et~al.},\ }\bibfield  {title} {\bibinfo {title} {Universal spin
  resonance in optimally doped antiferromagnetic superconductors},\ }\href@noop
  {} {\bibfield  {journal} {\bibinfo  {journal} {Phys. Rev. B}\ }\textbf
  {\bibinfo {volume} {81}},\ \bibinfo {pages} {064518} (\bibinfo {year}
  {2010})}\BibitemShut {NoStop}%
\bibitem [{\citenamefont {Sun}\ \emph {et~al.}(2023)\citenamefont {Sun} \emph
  {et~al.}}]{Sun2023Ni}%
  \BibitemOpen
  \bibfield  {author} {\bibinfo {author} {\bibfnamefont {H.}~\bibnamefont
  {Sun}} \emph {et~al.},\ }\bibfield  {title} {\bibinfo {title} {Signatures of
  superconductivity near 80 {K} in a nickelate under high pressure},\
  }\href@noop {} {\bibfield  {journal} {\bibinfo  {journal} {Nature}\ }\textbf
  {\bibinfo {volume} {621}},\ \bibinfo {pages} {493} (\bibinfo {year}
  {2023})}\BibitemShut {NoStop}%
\bibitem [{\citenamefont {Li}\ \emph {et~al.}(2024)\citenamefont {Li} \emph
  {et~al.}}]{Li2024Ni}%
  \BibitemOpen
  \bibfield  {author} {\bibinfo {author} {\bibfnamefont {Q.}~\bibnamefont {Li}}
  \emph {et~al.},\ }\bibfield  {title} {\bibinfo {title} {Signature of
  superconductivity in pressurized infinite-layer nickelates},\ }\href@noop {}
  {\bibfield  {journal} {\bibinfo  {journal} {Chin. Phys. Lett.}\ }\textbf
  {\bibinfo {volume} {41}},\ \bibinfo {pages} {017401} (\bibinfo {year}
  {2024})}\BibitemShut {NoStop}%
\end{thebibliography}
\end{document}